\documentstyle[epsfig,psfig,astrobib2,myaasmacros,mymacros,subfigure,lscape,multirow,amssymb]{mn2e}
\title[Star formation rate at $z\sim1$]
  {The star formation rate at redshift one: \Ha\ spectroscopy with CIRPASS}
\author[M.Doherty et al.]
 {Michelle Doherty$^{1,2}$\thanks{E-mail: mdoherty@eso.org},
Andrew Bunker$^{3}$,
Robert Sharp$^{4}$,
Gavin Dalton$^{5,6}$,
 \newauthor 
Ian Parry$^1$,
Ian Lewis$^5$\\
 $^1$Institute of Astronomy, Madingley Road, Cambridge, CB3~0HA, UK\\
 $^2$European Southern Observatory, Karl-Schwarzschild-Str 2.,Garching ,85748, Germany\\
 $^3$School of Physics, University of Exeter, Stocker Road, Exeter, UK\\
 $^4$Anglo-Australian Observatory, Epping, NSW, 1710, Australia\\
 $^5$Astrophysics, NAPL, Keble Road, Oxford, OX1~3RH, UK\\
 $^6$Rutherford Appleton Laboratory, Chilton, Didcot, OX110QX, UK\\ 
}
\date{Released 2005? Xxxxx XX}

\pagerange{\pageref{firstpage}--\pageref{lastpage}} \pubyear{2004}

\def\LaTeX{L\kern-.36em\raise.3ex\hbox{a}\kern-.15em
    T\kern-.1667em\lower.7ex\hbox{E}\kern-.125emX}

\def\Ha{\ifmmode \mathrm{H}{\alpha}\else H$\alpha$\fi}
\def\Msol{\ifmmode \mathrm{M}_{\sun}\else M$_{\sun}$\fi}

\begin{document}
 
\label{firstpage}

\maketitle

\begin{abstract}

We have conducted an \Ha\ survey of 38 $0.77 \leq z \leq 1$ galaxies over $\sim100$\,arcmin$^2$ of the Hubble Deep Field North and Flanking Fields,
to determine star formation rates (SFRs), with the near-infrared multi-object
spectrograph CIRPASS on the WHT. This represents the first successful application of this technique to
observing high redshift galaxies. Stacking the
spectra in the rest-frame to infer a total SFR for the field, we find a
lower limit (uncorrected for dust reddening) on the star
formation rate density at redshift $z=1$ of 0.04\Msol\,yr$^{-1}$\,Mpc$^{-3}$. This implies
rapid evolution in the star formation rate density from $z=0$ to
$z=1$ which is proportional to $(1+z)^{3.1}$.

\end{abstract}



\section{Introduction}

The quest for a global star formation history of the universe, as a key
element in understanding galaxy assembly, commenced around 10 years ago. Madau et al. (1996) combined the results from the
Canada-France Redshift Survey (CFRS) work \citep{llch+95,llhc96}, Steidel's
Lyman break galaxies (LBGs) at $z\approx3$ (Steidel et al. 1996a,b)\nocite{sgp+96}\nocite{sgda96} and a population of
faint galaxies with photometric redshifts $2<z<4.5$ in the Hubble Deep
Field (HDF), to give the first global view of the history of star formation
in the Universe. Diagrams in a similar vein were produced around the same
time by Lilly et al. (1996), at $z\lesssim1$ and
Pei \& Fall (1995\nocite{pf95}), based on damped Lyman alpha systems. 
Since those major works, there has been a volley of attempts to populate
the now famous `Madau-Lilly' diagram with data points at different epochs,
using observational studies in a variety of wavebands to infer the (volume-averaged) star formation history.

To date, many quantitative attempts to measure the global star formation history
have been based on optical measurements and have thus suffered from having to use different indicators of star
formation in various redshift bins, redshifted into the
optical. These various indicators not only have uncertain relative
calibration but are also affected differently by dust extinction. 
Commonly used star formation rate (SFR) indicators are ultra-violet (UV) continuum luminosity, which can be
heavily dust extincted, and nebular emission lines such as \Ha\ and [OII] (the
latter of which is strongly dependent on metallicity and ionisation state).
A thorough review of these different techniques and their relative calibrations is given by Kennicutt
(1998\nocite{ken98}). 

Longer wavelength estimators relying on Far-infrared (FIR) or radio
luminosity are insensitive to dust obscuration yet have their own
caveats. For example, radio luminosity can also be produced by AGN, therefore this method requires cross-correlating
with optical spectroscopic features to eliminate AGN from the sample. Sub-millimetre
estimates rely heavily on assumptions about the dust temperature (e.g. Hughes et
al. 1998\nocite{hsd+98}) and a significant fraction of sub-mm galaxies also
turn out to harbour AGN \citep{cbsi05}. 

The Universe at redshift $z\sim1-2$ is believed to be one of the most active epochs
in galaxy formation and evolution. Indeed, it is inferred to be the epoch at which large elliptical and spiral galaxies are assembled  and therefore may also be the period of peak star formation in the Universe. Yet the Universe at this epoch is still not well studied nor understood. Observations have long been hampered by the difficulties of observing objects at these redshifts in the visible wavebands.
At redshift $z\sim1$, key diagnostic spectral features are redshifted out
of the optical into the near-infrared, which is a difficult regime to work
in, and the rest-UV Lyman$-\alpha$ line is not accessible in the optical
until around redshifts $z\sim2.5$. The redshift range $z\sim1-2$ has hence
traditionally been dubbed the spectroscopic `redshift desert', although there has been
some recent progress in exploring this epoch through extending colour-selection techniques to star-forming galaxies at 
$z\approx1.5-2$ (e.g. Adelberger et al. 2004\nocite{ass+04} , Steidel et al. 2004\nocite{ssp+04}). The Gemini Deep Deep Survey \citep{agm+04} and the K20 survey \citep{cdm+02} have also identified some passively evolving galaxies at these epochs. 

There is certainly evidence that the star
formation rate was much higher in the recent past ($z\gtrsim0.5$), compared with the
current epoch (e.g. Lilly et al. 1996,
Tresse et al.  2002, Hippelein et
al. 2003)\nocite{llhc96}\nocite{tmlc02}\nocite{hmm+03}. 
However, it is still unclear whether at redshifts of one and beyond the star formation density
plateaus, or declines or perhaps continues to increase.
 
Even in the most up-to-date versions of the Madau-Lilly diagram (e.g. Hopkins
  2004 \nocite{hop04}) there is still almost an order of magnitude discrepancy between different indicators, at $z\approx1$. In order to obtain a self-consistent global picture of the SF history of the Universe, the same robust indicator needs to be employed across all redshift bins. Researchers have begun to converge on the \Ha\ emission
line as such a robust indicator. It is a good tracer of the {\em
  instantaneous} SFR as it is directly proportional to the ionising UV
  Lyman continuum radiation from the most short-lived ($t_{MS}<20$\,Myr) and
 massive ($>10\Msol$) stars.
\Ha\ has been widely used in surveys at low redshift (e.g. Gallego et
  al. 1995). It is particularly
suitable as it is relatively immune to metallicity effects and is much
less susceptible to extinction by dust than the rest-UV continuum and
Lyman-$\alpha$ (which is also selectively quenched through resonant
scattering). 
The calibration required to derive SFRs from \Ha\ assumes that no Lyman continuum photons
escape the galaxy, which is almost true -- at most the escape fraction may be
around 5\% (see discussion in Glazebrook et al.\ 1999 and references
therein).

 Although the models also depend on star formation history, galaxy age and
 metallicities, it is the IMF to which the SFR calibration is most
 sensitive as the \Ha\ emission is produced almost entirely by the most
 massive stars. The Kennicutt (1998) conversion used in this paper is calculated for Case B recombination at $
T_e=10,000$K (e.g. Osterbrock 1989\nocite{ost89}) and assumes solar
 metallicities and
 a Salpeter (1955\nocite{sal55}) IMF\footnote{A Scalo (1986) IMF
yields star formation rates a factor of three times higher for the same
 \Ha\ flux.} with mass cut-offs 0.1 and 100\Msol . The conversion is given by:
\begin{equation}
{\rm SFR}(\Msol~{\rm yr}^{-1})=7.9\times10^{-42}~L(\Ha )~~~~~(\ergs)
\label{eqn:K98}
\end{equation}

At $z\sim1$ \Ha\ is shifted into the near-infrared and until recently
near-infrared spectroscopy has been restricted to long-slit work and
building samples using single object spectroscopy is inefficient in
terms of telescope time. Furthermore, the small statistical samples obtained result
in large uncertainties in the global properties of galaxies at $z\sim 1$. 
For example, Glazebrook et al. (1999\nocite{gbe+99}) targeted a sample of
13 $z\sim1$ field galaxies in the CFRS and found star formation rates 2--3
times higher than those inferred from the UV continuum. Tresse et
al. (2002)\nocite{tmlc02} have a larger \Ha\ sample of 33 field galaxies at
$0.5<z<1.1$ and have built the \Ha\ luminosity function. Comparing this to their lower redshift sample (Tresse \&
Maddox 1998\nocite{tm98}) and other work in the literature, they find a strong rise in star formation rate
-- a factor $\sim12$ from $z=0.2$ to $z=1.3$.  Erb et al. (2003\nocite{ess+03}) obtained a small
sample ($\sim16$ galaxies) of \Ha\ measurements at $z\sim2$ for the
purposes of dynamical
rotation curves but did not derive the SFR density. 

There has been successful `multi-object' grism spectroscopy through slitless surveys in the near-infrared from space, but such an approach has poor sensitivity because of the high near-infrared background due to zodiacal light. 
 In the HST/NICMOS survey of McCarthy et al. (1999\nocite{myf+99}) 33
 emission line galaxies were identified between $0.7<z<1.9$. From this data
 set Yan et al. (1999)\nocite{ymf+99} derived the \Ha\ luminosity function and therefore the SFR density, again finding a result which is an order of magnitude higher than the local Universe. Hopkins et al. (2000\nocite{hcs00}) repeated this experiment with an independent data set of 37 \Ha\ emitting galaxies in $\sim4.4$\,arcmin$^2$, over a redshift range $z=0.7-1.8$ and found a consistent result.

It is only now that true multi-object infrared spectroscopy is possible from the ground, using
either a slitmask approach (e.g. IRIS\,2 on AAT and FLAMINGOS on Gemini and
KPNO) or a fibre-fed spectrograph. 

We have used the Cambridge InfraRed
PAnoramic Survey Spectrograph (CIRPASS; Parry et al. 2000) \nocite{pmj+00} in multi-object mode 
with the aim of addressing the true star formation history of the Universe at redshifts
$z=0.7-1.5$, through \Ha\ measurements of a large sample of galaxies.

 CIRPASS can operate with 150 fibres with the ability to simultaneously
 observe up to 75 targets (in object/sky pairs) and demonstrates a
 powerful new technique for studying distant galaxies. 
Initial results from
this project have been presented in Doherty et al. 2004. This is the first successful example of near-infrared
 multi-object spectroscopy of high redshift galaxies. Here we present
 results from this data set on the statistical measurement of the volume
 emissivity of \Ha\ at redshift one, and hence the total star formation
 rate density of the universe at this epoch. 

The format of the paper is as follows. In Section 2 we give a brief overview of the instrument and the observations. A full description of the instrument set-up and data reduction is given in Doherty et al. 2004\nocite{dbs+04}. 
We discuss the sample selection and completeness in some detail in Section
3.
Section 4 covers the procedures behind stacking the spectra and
calculating the total star formation rate of our sample and in Section 5 we
discuss the necessary corrections to estimate a total volume-averaged SFR at redshift one, from the
combined SFR of our sample.
The reader familiar with this field and interested in our result in the context of evolution in the SFR density over cosmic time, may wish to skip directly to Section 6,
where we compare our measurement to other \Ha\ results from $z=0-2$.

In this paper we adopt the standard ``concordance''
cosmology of $\Omega_M=0.3$, $\Omega_{\Lambda}=0.7$, and use
$h_{70}=H_0/70\,{\rm km\,s^{-1}\,Mpc^{-1}}$. AB magnitudes \citep{og83}
are used throughout.

\section{Observations and data reduction}

CIRPASS is a near-IR fibre-fed spectrograph operating
between 0.9 and 1.67$\mu$m. The upper cut-off is set by a blocking filter
which reduces the thermal background. CIRPASS can operate in one of two modes -- with an
Integral Field Unit or in multi-object mode \citep{pmj+00}.
The CIRPASS Multi-Object Spectrograph (CIRPASS-MOS) was used at the Cassegrain focus of the
 4.2m WHT in La Palma, to observe 62 objects at a time in the Hubble Deep
Field North (HDFN; \citet{wbd+96}). 
The fibre size corresponds to $\sim$1.1\arcsec\ at the WHT (which is approximately comparable to the expected
seeing convolved with typical galaxy profiles, at least for compact
galaxies). However, there are some galaxies in our sample which are extended and
closer to $\sim$2\arcsec\ across (as evident from HST imaging in the HDFN) and for these some light is therefore lost. At the WHT the
fibres are deployable over an unvignetted field of 15\arcmin\ diameter, with
a minimum separation of 12.6\arcsec .
A Hawaii 2K detector was used and a grating of 831 lines/mm,
producing a dispersion of 0.95\AA\,pixel$^{-1}$. The FWHM of each fibre
extends over 2.7 pixels in both the spatial and spectral domain. The
wavelength coverage with the 2K detector was 1726\AA, covering most of the $J-$band, and the
grating was tilted to set a central wavelength of $\lambda_c \approx 1.2
\mu$m. We acquired the field using
six guide bundles, each containing seven closely packed fibres, centred on
bright stars. We thus positioned
the plate to an accuracy better than 0.5\arcsec, that is, less than
half the fibre size.

 The observations were taken with the CIRPASS-MOS on the WHT, spanning a
 total of 6.5hours over two nights. These are summarised in Table~\ref{tab:obs1}.
\begin{table*}
\caption{Observations in the HDF-N. The data were taken over two nights with a total of 6.5 hours exposure time. The telescope was nodded by a distance 6.08 \arcmin (beamswitch offset) between the object and sky fibres. }
\begin{tabular}{cccccccccp{3cm}}
\hline
\hline
Field & field & Date &  Exposures &  beamswitch & $\lambda _c$ &no. & data set\\
name  & centre &     &            &  offset       &   (\AA)    & objects  & \\
      & (J2000) &    &            &         &              &       &  \\
\hline
HDFN & 12 36 25.63,  & 27/12/03 &  7$\times$1800s     & 6.08 \arcmin &12217 & 62 & Night I\\
HDFN &  +62 13 46.5  & 02/01/04 &  2$\times$3000s +2$\times$2400s  & 6.08 \arcmin &12217 & 62 & Night II\\
 \hline

\end{tabular}
\label{tab:obs1}
\normalsize
\end{table*}
Precise details of the instrument setup and data reduction process are given in Doherty et
 al. (2004).

Flux calibration was carried out to account for both the spectral response of the instrument and the atmospheric transparency. Observations of a standard star were used to calibrate
the shape of the spectral response function (normalised to one in the
middle of the detector) and the atmospheric absorption. The flux
normalisation was then obtained from a bright 2MASS star ($J=10.68$)
 which was observed simultaneously with the targets
i.e. with the same fibre plug plate.
 As the star and objects were observed simultaneously, this method corrects
for temporal changes in the seeing, which fluctuated considerably over the
course of the observations. A second, fainter ($J=13.247$) star
was also observed but may have been vignetted as it was close to the edge
of the plate and the relative fluxes are consistent to only
30\%. Furthermore, the line fluxes of our detected objects vary between the two nights within this 30\% error margin. This might suggest that there was a small rotational error in aligning the plate and we can thus only guarantee the flux calibration to within 30\%.

The gain of the 2K detector was measured to be 4.5\,$e^{-}\,{\rm DN}^{-1}$ where DN is data numbers, or counts.
We kept track throughout the reduction process of the number electrons in
the background noise, for Poisson statistics used in calculating the noise
and hence the error on the final SFR (see
Section~\ref{subsec:codefluxes}). We are in a Poisson regime as multiple
readouts were taken to reduce the readnoise to a negligible level.

\section{Sample selection and completeness}
\label{sec:complete}
Our sample was selected from \cite{chb+00} -- a very complete, magnitude limited,
spectroscopic follow-up of the HDF-N. This catalogue also includes the `Flanking Fields' (FF) -- an approximately
circular area of radius 8\arcmin\ centred on the HDF-N and imaged with WFPC2
for $\sim$2 orbits in a single passband and the size of which is well
matched to the CIRPASS-MOS field of view.

The Cohen et al. (2000) catalogue is $>90\%$
complete to $R<23$ in the flanking fields and $R<24$ in the HDF-N proper (AB magnitudes). 
In the catalogue there are 157 galaxies in our range $0.7<z<1.0 $, but we
selected only galaxies identified with emission lines (i.e. excluding
absorption line systems classified `A'), giving 137. These `A' systems not
only make up a small fraction (13\%) of all targets in our redshift
range but are likely to be either post-starburst or highly evolved galaxies
with no current star formation. We therefore would not expect to detect
\Ha\ in these and chose not to allocate fibres to them, as they are
unlikely to contribute much to the SFR density. 

We observed a random sub-sample of 62 of those 137
galaxies, with perhaps only a small bias against close galaxy pairs, as the
physical size of the fibres meant they had to be placed at least
12.6\arcsec\ apart on the sky. Where there was a position clash within that
distance, the brighter source in $R-$magnitude was given priority. 

Sources which have redshifts placing \Ha\ in the atmospheric absorption
trough in the $J-$band at $1.10-1.16\mu$m were excluded from the sample {\it a posteriori}, as
strong atmospheric features create spurious signals in the stacked spectrum
(Section~\ref{sec:stack}). This region is at the short-wavelength end of the
array and taking this cut effectively increases our minimum redshift to
$z=0.768$, decreasing the surveyed volume but increasing the confidence in
the stacked detection. 

Our observed target list includes a small number of galaxies in magnitude
bins where the catalogue completeness is low (i.e $R>23.5$ in the FF and
$R>24.0$ in HDF). In order to derive a meaningful statistical result for
the star formation in the field, we need a consistent sample to some
limiting magnitude. We therefore take a cut at $R=23.5$ where there is $\sim$84\% completeness in the catalogue (with most of
the incompleteness in the last 0.5 magnitude bin in the FFs).

In summary, the sample adopted for the purposes of a statistical measurement of the SFR
is restricted to $0.768<z<1.0$ and $R<23.5$ and numbers 38 objects (out of 91 possible targets for observation).  We therefore adjust the results by dividing by this incompleteness factor (38/91)$=$0.42. 

Guide stars
were taken from the Guide Star Catalogue-II (Morrison et
al. 2001\nocite{mm+01}).
 For consistency in the astrometry between our targets and
alignment stars, we redetermined the coordinates of each target object using the {\sc GOODS}\,v1.0
images (Giavalisco \& GOODS Team 2003)\nocite{gia03}. The typical offset from the Cohen et al. (2000) coordinates was 0.5\arcsec\ (note that for a few extended galaxies our centre is also slightly different from the position listed
in the GOODS catalogue).

\begin{figure}
\begin{tabular}{c}
\psfig{figure=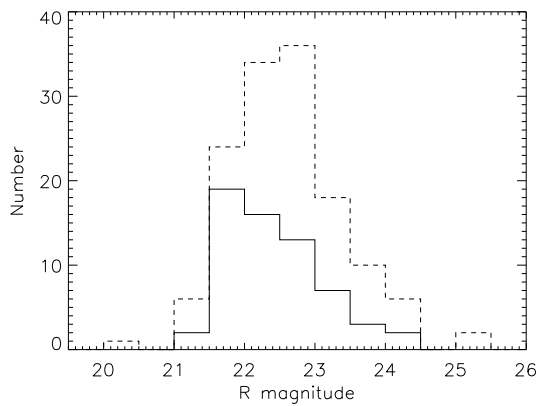,angle=90,width=8cm} \\
\psfig{figure=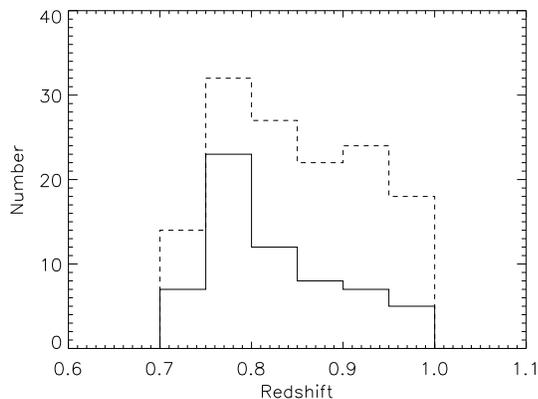,angle=90,width=8cm} \\
\end{tabular}
\caption[Histograms in R-magnitude and redshift for the HDF sample]{\small{Histograms in R-magnitude (top) and Redshift (bottom) for all objects in the Cohen et al. (2000) catalogue in our redshift range (dashed line) and our observed sample (solid line). There seems no significant bias in magnitude or redshift space, particularly for our restricted sample with $0.768<z<1.0$ and $R<23.5$.}}
\label{fig:complete_histos}
\end{figure}

As a test for any bias in redshift or magnitude space we plot the
histograms for our observed sub-sample versus all of the possible targets
from the catalogue in our survey volume (Figure~\ref{fig:complete_histos}). There seems no significant bias in magnitude or redshift space, particularly for our restricted sample with $0.768<z<1.0$ and $R<23.5$.

\begin{figure}
\centering
\psfig{figure=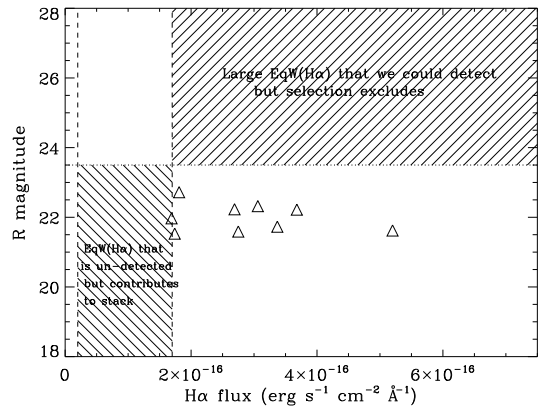,angle=90,width=8cm} \\
\caption[Equivalent width problem. \Ha\ v's $R-$magnitude diagram.]{\small{\Ha\ v's $R-$magnitude diagram. The shaded regions show the high EqW(\Ha) sources we would miss (in the \Ha
  $-R$-mag plane) by taking a cut of $R=23.5$ and the sources which we don't detect, but which contribute to the stacked flux (i.e. the statistical measure of the SFR, see Section~\ref{sec:stack}). The sources with $>5\sigma$ \Ha\ detections are plotted as triangles.}}
\label{fig:HaRmag}
\end{figure}

Finally, there is a source of incompleteness introduced by taking a broad-band limit for the selection. Sources with large \Ha\ equivalent widths 
(EqW(\Ha)), i.e. sources which are luminous
in \Ha\ but too faint in the $R-$band to make the $R<23.5$ magnitude cut (Figure~\ref{fig:HaRmag}), may be missed.
We might reasonably expect there to be a population of sources with faint continuum magnitudes but
large \Ha\ flux, perhaps young, actively star forming galaxies with minimal evolved populations
(which tend to dominate the longer wavelengths). Such systems
would be missed in surveys like this which rely on magnitude-limited
catalogues to pre-select targets with known spectroscopic redshifts. 

However, narrow-band imaging searches which select directly on the emission
line flux  (in this case \Ha) find few candidates and would suggest that
the hypothetical population proposed above is extremely rare (if it exists at all). For example, Willis \& Courbin (2005)\nocite{wc05} carried out a deep
narrow-band search in the $J-$band, targeted at Ly$-\alpha$ at $z\sim9$ to a
limiting flux of $3\times10^{-18}$\ergscm, over $2.5\times2.5$\,arcmin$^2$. The search is centred at
1.187$\mu$m and one might expect the by-product of \Ha\ detections at
$z=0.8$. In fact they find only two sources with a narrow-band excess, one
of which is likely \Ha\ and the other H$\beta$.   

Glazebrook et al. (2004)\nocite{gtt+04} used a tunable filter with precisely the aim of searching for
line-emitters which would be missed in surveys such as ours. They observed
three slices at 7100, 8100 and 9100\AA\ (covering [OII], H$\beta$, \Ha\ over the redshift range 0.3--0.9), to a flux limit of $2\times10^{-17}$ \ergscm . They find no evidence for any new strongly line-emitting population of low luminosity objects and find an SFR density consistent with estimates from broad-band selected samples.  

For a typical detected galaxy ($R_{AB}\approx 22$; Doherty et al. 2004) in our \Ha\ survey, the limiting equivalent width at our $5\,\sigma$ flux threshold
of $1.9\times 10^{-19}\,{\rm erg\,cm^{-2}\,s}$ is $EW_{\rm rest}>35$\,\AA , assuming a colour of $(R-J)_{AB}\approx 1$
typical of star-forming galaxies at $z\approx 1$.  For the broadband selection limit $R_{AB}=23.5$, an EW closer to 140 \AA\  would be needed to detect \Ha\  emission above our limiting flux.

\section{Star Formation Rate from composite spectrum}
\label{sec:stack}
Confident that we have observed a random sub-sample of the whole sample (as
discussed in Section~\ref{sec:complete}), we now stack the spectra in rest-frame wavelength to obtain statistically a firm lower limit on the  total star
formation rate per unit volume in our surveyed field. We subsequently discuss various corrections that can be made to estimate the true star formation rate density (e.g. incompleteness due to luminosity bias, aperture corrections and reddening corrections) and finally place this measurement in the context of evolution from $z\sim2.5$ to $z\sim 0$.

\subsection{Stacked spectrum}

Any remaining skyline residuals due to imperfect background subtraction that are
close in wavelength to the \Ha\ line could skew a statistical
measurement of \Ha\ (although note that none of the Cohen redshifts place
\Ha\ within 4\AA\ of a skyline). We therefore firstly mask out and interpolate over
skyline residuals, where the sky is brighter than some threshold value, in the observed frame. 

Using the \cite{chb+00} redshifts, which are mostly determined from
[OII], each spectrum is shifted to the rest-frame, interpolated onto a common
wavelength grid, which is over-sampled to 0.5\AA\ per pixel and summed.

\begin{figure}
\centering
\begin{tabular}{c}
\epsfig{figure=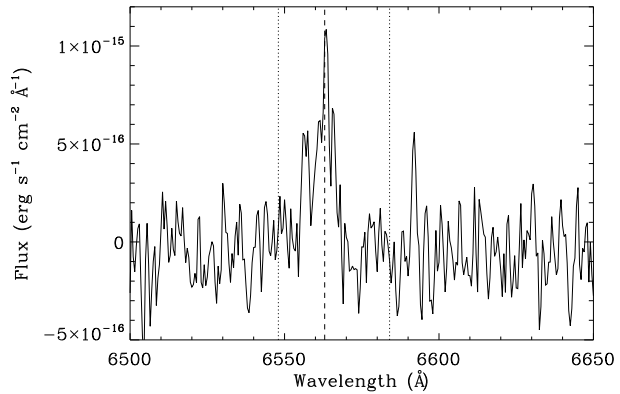,angle=90,width=0.45\textwidth}\\
\small{\bf{(a) NIGHT I}}\\
\epsfig{figure=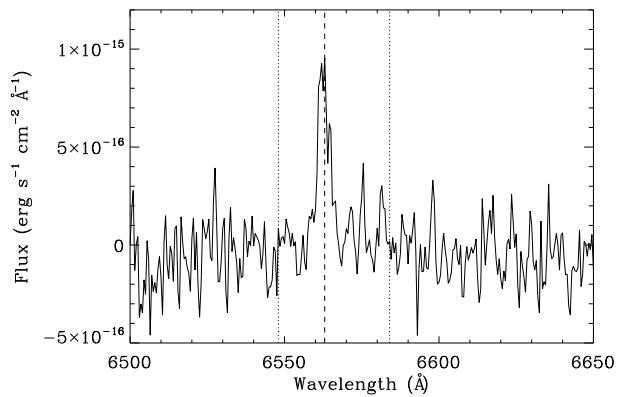,angle=90,width=0.45\textwidth}\\
\small{\bf{(b) NIGHT II}}\\
\end{tabular}
\caption{\small{Stacked spectrum for each night's observations in the HDF-N, excluding sources where \Ha\
        lands in the atmospheric trough and restricting to $R\leq23.5$. 38 sources in total. The central wavelength of \Ha\ (6563\AA ; from optical redshifts) is marked with a dashed line. The expected positions of [NII] are marked with dotted lines, but they are undetected.} }

\label{fig:stack}
\end{figure}

The stacked spectra (using a straight sum with no weighting) are shown in
Figure~\ref{fig:stack}. The \Ha\ signal in the stacked spectrum from the
Night I data is broader than that of Night II, in particular due
to the presence of a feature just blue-ward of the \Ha\ line (around
$\sim$6555\AA). It is possible that it may be due to additional
\Ha, perhaps from sources with slightly incorrect optical redshifts. We
visually inspected the individual spectra contributing to the stack and
investigated stacking them cumulatively, to ascertain whether this
feature represents a real signal. The sky subtraction for this night was
not as effective as Night II, and from inspection we deduce that this may be an accumulation of several sky residuals (in spite of
all efforts to  minimise the presence of skylines). Furthermore, since
there is no evidence for a repetition of this signal in Night II data
(where the sky subtracted out in a much cleaner fashion), we
conclude that it is spurious. 

There is no evidence of [NII] in the stacked spectrum. However, the [NII] emission
is known to have a systemic velocity offset from \Ha , which may have a
smearing effect in the stack. 

\begin{figure}
\centering
\begin{tabular}{c}
\epsfig{figure=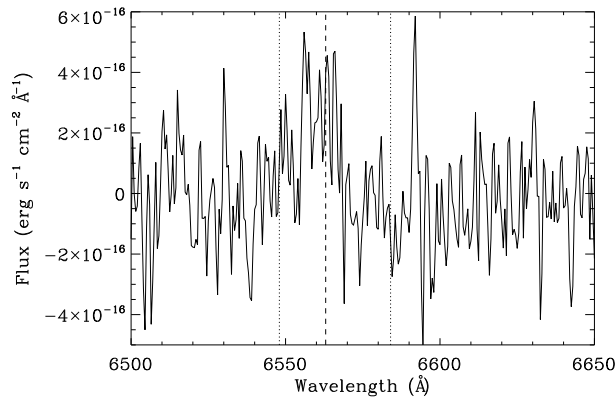,angle=90,width=0.45\textwidth}\\
\small{\bf{(a) NIGHT I}}\\
\epsfig{figure=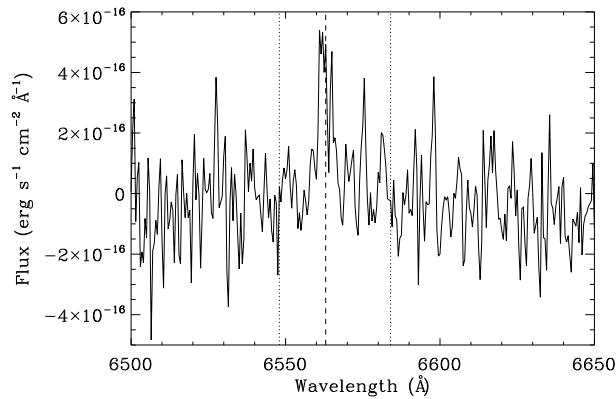,angle=90,width=0.45\textwidth}\\
\small{\bf{(b) NIGHT II}}\\
\end{tabular}
\caption{\small{Stacks for each night excluding those sources detected at $>5\sigma$. 29 sources total.}}
\label{fig:nondetections}
\end{figure}

If we exclude the objects with robustly detected \Ha\ ($>5\sigma$) from the
stack, there is still a significant combined \Ha\ contribution from the
nominally undetected objects -- around half of the flux in the total stacked
spectrum (Figure~\ref{fig:nondetections}).

\subsection{Calculating the total stacked flux and luminosity}
\label{subsec:codefluxes}

For each object we measure the \Ha\ flux and hence line luminosity and SFR,
given its redshift. The sum of these SFRs represents the total star
formation rate for our sample of galaxies.

\begin{figure}
\centering
\epsfig{figure=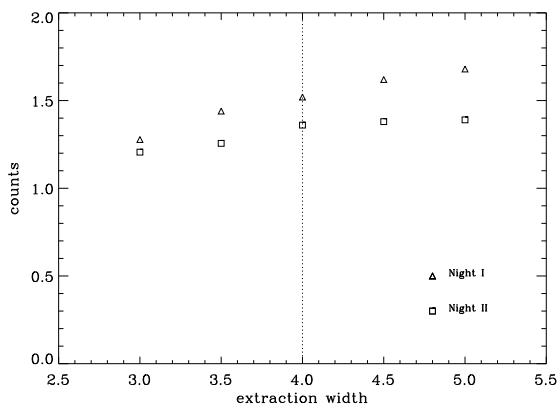,width=8cm}
\caption[\Ha\ counts v's wavelength extraction width]{\small{Total \Ha\ flux in the stack (in counts), as a function of the extraction half-width taken.}}
\label{fig:ext_radius}
\end{figure}

Figure~\ref{fig:ext_radius} shows the curve of growth for the total \Ha\
flux  for each night's {\it stacked} spectrum, as a function of the wavelength extraction
half-width (in the rest-frame). It is evident that beyond a half-width of
4\AA, the summed flux from night II is rapidly converging, whilst that of
night I continues to rise, due to the gradual inclusion of the noise spike
corresponding to the spurious feature seen in the stacked spectrum. 
We therefore determined the \Ha\ line flux for each object by
integrating over a wavelength region of $6563\pm4$\AA\ rest-frame
(i.e. within $\sim200$\kms\ of the [OII] redshift). This
extraction width optimises the real signal included in the final sum whilst
excluding additional noise. The average baseline width between zero power points for the $>5\sigma$
detections is 13\AA\ in the observed frame (see example spectra in Doherty et al. 2004, Figure~1). Our extraction width used here is slightly larger and therefore encompasses the small velocity
offsets between the [OII] and \Ha\ emission.

To combine the data from the two nights, which should improve the S/N,  the weighted mean of the two luminosity measurements for
each object is taken according to:
\begin{equation}
\bar{X} = \frac{\sum_{i=1}^n X_i/\sigma_i^2}{\sum_{i=1}^n 1/\sigma_i^2}
\end{equation}
where $X_i$ are independent estimates of the data point with associated
errors $\sigma_i$,
and the variance is given by:
\begin{equation}
\sigma^2 = \frac{1}{\sum_{i=1}^n 1/\sigma_i^2}.
\end{equation}
This method effectively reduces the error, providing a cleaner measurement of the total \Ha\ flux. 
The noise in the integrated flux for each spectrum is calculated using Poisson statistics from the
background noise, rather than measuring the standard deviation between
pixels in the final spectra, as the noise is correlated after repeated
extractions, interpolations and rebinnings. 

The conversion to total star formation rate for the sample is then given by
Kennicutt's (1998)\nocite{ken98} calibration (Equation~\ref{eqn:K98}).

 The fluxes, luminosities and SFRs are shown in
 Table~\ref{tab:fluxes}. The resulting combined SFR is a factor of $\sim2$
 higher than the sum of the individual $>5\sigma$ detections. There is
 clearly a significant contribution to the total \Ha\ flux from objects
 which are nominally undetected in the individual spectra (Figure~\ref{fig:nondetections}). 
This therefore becomes a powerful technique for obtaining the total SFR in
a surveyed volume, particularly when one can quantify the sample
completeness.

\begin{table*}
\footnotesize
\hspace{-2cm}
\caption[\Ha\ fluxes, luminosities and SFRs for the stacked spectra]{\small{\Ha\ fluxes, luminosities and SFRs for each nights' data. Measurements are listed for different subsets of the sample: i) all objects contributing to the stack, i.e. 38 objects with $0.768<z<1.0$ and $R<23.5$, ii) $>5\sigma$ detections only (9 objects), iii) sub-sample excluding the $>5\sigma$ detections (29 objects).}}
\begin{tabular}{|c|c|c|c|c|c|c|}
\hline
\multicolumn{7}{c}{{\bf Night I}}\\
\hline
Sub- set&\#&\Ha\ flux       &S/N&\Ha\ flux          &$L_{\Ha}$        &SFR    \\
        &  &counts\, s$^{-1}$&   &($\times10^{-15}$)&($\times10^{42}$)&       \\
        &  &                &   &\ergscm \AA $^{-1}$&  \ergs          &\Msol /yr\\
\hline  
i)&  38 & 1.56$\pm$0.07 & 21.4 & 5.07$\pm$0.26 & 18.28$\pm$0.86      &144.4$\pm$6.8 \\
ii)&  9 & 0.79$\pm$0.03 & 24.8 & 2.61$\pm$0.11 & 8.94$\pm$0.36       &70.6$\pm$2.9  \\
iii)&29 & 0.77$\pm$0.07 & 11.7 & 2.46$\pm$0.23 & 9.34$\pm$0.78       &73.8$\pm$6.2  \\
\hline 
\multicolumn{7}{c}{{\bf Night II}}\\
\hline 
i) &38 & 1.47 $\pm$ 0.08 & 17.5 & 4.29 $\pm$ 0.25 & 14.41$\pm$ 0.84 & 113.8$\pm$  6.7  \\
ii)  & 9 & 0.67 $\pm$ 0.04 & 16.3 & 1.90 $\pm$ 0.12 & 6.46 $\pm$ 0.39 & 51.1 $\pm$ 3.1 \\
iii) &29 & 0.8  $\pm$ 0.07 & 10.9 & 2.39 $\pm$ 0.22 & 7.95 $\pm$0.75  & 62.8 $\pm$ 5.9 \\
\hline
\end{tabular}
\label{tab:fluxes}
\end{table*}

\section{Global star formation rate density}

The SFR quoted above ($131\pm39 \msol$\,yr$^{-1}$)
is a firm lower limit on the true SFR for the total sample, as only a
subset were spectroscopically targeted. In the following sub-sections we
apply a series of corrections, leading to an estimate which is closer to
the true SFR for the field, and hence derive the global star formation rate
density (SFRD) at $z\sim1$. We then compare this value to other \Ha\ measurements of the SFRD in the context of evolution from a redshift of around $z\sim2.5$ to the present-day.

The first, simple, correction is to account for the incompleteness due  to
crowding (i.e. the fact that we did not observe all galaxies in the field
down to the limiting magnitude, but only a random subsample which could fit
on the plug-plate without fibre clashes). This was discussed in detail in
Section~\ref{sec:complete} and amounts to dividing the results by 0.42,
giving a corrected SFR of $312\pm93\Msol\,yr^{-1}$. 
The remaining corrections rely on various assumptions, which shall be clearly stated and discussed as we proceed. The steps, briefly, are as follows:
\renewcommand{\theenumi}{\arabic{enumi})}
\begin{enumerate}
\item {\bf Incompleteness correction} for galaxies below the survey flux
  limit. This is an effect dependent on the inherent luminosity function --
  that is, we do not probe the entire survey volume down to the same
  luminosity limit. In reality we have an approximately uniform flux limit
  (determined by the sensitivity of the instrument) which means that the
  luminosity limit drops off as  $1/d_L^2$ where $d_L$ is the luminosity distance. In other words, more distant galaxies must be intrinsically more luminous in \Ha\ in order to be detected within our survey. To account for this incompleteness and estimate the true SFR density, it is necessary to make some assumptions about the \Ha\ luminosity function in our redshift range.
\item {\bf Aperture corrections}. The fibre size on the sky is only
  1.1\arcsec\ in diameter, which is around the size of the compact
  galaxies, convolved with the seeing (seen in HST imaging, e.g.
  Figure~2 in Doherty et al. 2004). However, for more extended galaxies
  we may be missing a substantial fraction of the \Ha\ flux, particularly
  if there are localised star-forming regions which fall outside the fibre
  aperture, which indeed appears to be the case, at least for some of the
  galaxies in the sample (see, for example, objects a), c) f) in
  Figure~2 of Doherty et al. 2004). There is no way of
  ascertaining aperture corrections for the \Ha\ flux directly from our
  data set, instead we will attempt to infer aperture corrections based on $B-$band photometry (corresponding to the rest-frame UV).
\item {\bf Reddening correction}. As \Ha\ is nominally undetected in most
  of our targets, it is not possible to derive actual reddening values for
  all sources. From the $B-$band luminosities measured for aperture
  corrections we determine the SFR$_{UV}$ and hence using the average SFR(\Ha)/SFR(UV) ratio, derive an average reddening correction
  using two different extinction laws -- a Milky Way type law and the
  Calzetti law. This is a useful exercise to place our SFRD measurement in context with other work -- many authors have applied reddening corrections to their data before deriving SFRDs.  

\end{enumerate}

\subsection{Incompleteness and the Galaxy Luminosity Function}
\label{sec:lumbias}
The luminosity distribution of galaxies in the Universe is known to be well-characterised by a Schechter (1976)\nocite{sch76} function, in which the number of galaxies per unit volume, $\phi (L)dL$, in the luminosity bin $L$:$L+dL$, is given by: 
\begin{equation}
\phi (L) dL = \phi ^*(L/L^*)^{\alpha}{\rm exp}(-L/L^*)d(L/L^*)
\label{eqn:schechterfunc}
\end{equation}
This function is parametrised by $\phi ^*$, $L^*$ and $\alpha$, where
\phistar\ is a number per unit volume, \Lstar\ defines the `knee' of the
function, and $\alpha$ gives the `faint end slope', the slope of the luminosity function in the log($\phi$)$-$log($L$) in the region $L\ll\Lstar$. Equation~\ref{eqn:schechterfunc} integrates to an incomplete gamma function, which can be solved numerically to find the expected number density $n$ of galaxies brighter than a certain luminosity limit $L_{\rm lim}$:
\begin{equation}
n(L>L_{\rm lim}) = \int_{x_{\rm lim}\Lstar}^{\infty} \phi (L) dL = \phistar \Gamma (\alpha + 1, x)
\label{eqn:numdensity}
\end{equation}
and the total luminosity in those galaxies:
\begin{equation}
L(L>L_{lim})=\phistar \Lstar \Gamma (\alpha +2,x)
\label{eqn:total-lum}
\end{equation} where  $x=L/\Lstar $. 
For a faint end slope $-2<\alpha <-1$, the function diverges for the number counts but not the luminosity (i.e. there can be an infinite number galaxies but not infinite luminosity -- many very low luminosity galaxies contribute almost nothing to the total luminosity). However, for a faint end slope $\alpha <-2$ there would also be infinite luminosity. 
  
Knowing the luminosity function to some degree of accuracy allows us to
correct for the incompleteness in surveys introduced by the
luminosity--distance dependence. In practice, with only 9 solid detections,
we do not have enough data points to construct a credible \Ha\ luminosity
function. However, we can attempt to model the incompleteness by
investigating a range of \Ha\ luminosity functions based on previous
work\footnote{converted to our $\Lambda$CDM cosmology. See Appendix
  for the method behind this.}, comparing our observed result with the predicted number counts and SFRs of star-forming galaxies which we should expect to see above the flux limit, over our redshift range.

Given various input luminosity functions, parametrised by \Lstar , \phistar and $\alpha$,
we loop over redshift and over luminosity and for each redshift slice
compute the expected numbers and SFRs in each luminosity bin (using
equations \ref{eqn:numdensity} and \ref{eqn:total-lum}), thereby producing a predicted
number density as a function of limiting line flux. We compare to our 5$\sigma$ flux limit in a single average fibre, $f_{\Ha}=1.9\times10^{-16}\ergscm $ and to the effective 5$\sigma$ flux limit
obtained by stacking up the 38 observed galaxies in our survey field,
$f_{\Ha}=3.3\times10^{-17}\ergscm $.  Table~\ref{tab:lumlimits} gives a summary of the result of this investigation. It lists the input luminosity function parameters, the corresponding predicted number counts and SFRs we should expect to observe in our surveyed volume at the above flux limits, and the total SFR (obtained by integrating down the luminosity function). This method allows us to infer what order of correction is needed to estimate the total SFR for our field. 

Yan et al. (1999) \nocite{ymf+99} derived an \Ha\ luminosity function for
galaxies at $0.7<z<1.9$ using slitless spectroscopy with the near-infrared
camera and multi-object spectrometer (NICMOS) on the HST. In that survey
they detect 33 emission line galaxies over 64\,arcmin$^2$
\citep{myf+99}. However, we note that this number seems very low if their
luminosity function is correct (see Table~\ref{tab:lumlimits} where the
predicted number counts are high). Their average 3$\sigma$ line flux limit
is $4.1\times10^{-17}\ergscm$ \citep{myf+99}.  Using their derived
luminosity function (LF) to model the galaxy counts and SFR we would expect to see in our survey volume, we find that either the $L^*$ or $\phi^*$ is significantly too large to account for our observed numbers -- it predicts around 3--4 times the number that we actually detect (Table~\ref{tab:lumlimits}).  The large value of $L^*$ implies that the majority of the star formation is in galaxies brighter than our $5\sigma$ limit, meaning if their luminosity function is accurate and appropriate for $z\sim1$ galaxies, we should have detected most of the objects we observed in \Ha. Hopkins et al. (2000)\nocite{hcs00} carried out a similar experiment with NICMOS grism spectroscopy and find a similar star formation rate density, albeit a very differently shaped luminosity function. 
It is difficult to reconcile our result with those works if we take their measurement to be representative of the redshift $z\approx1$ Universe.

Turning now to compare our results to the work of
\cite{tmlc02}\footnote{Tresse et al. (2002) have applied reddening
  corrections to their data before fitting a luminosity function. However,
  they cite an average correction of $A_v=1$ mag (a factor of 2.02 in \Ha\ luminosity) and we use this value to
  artificially `redden' their LF for the sake of comparison.}, who derive an LF from
$\sim30$ emission line galaxies at $0.5<z<1.1$ in the CFRS, we find a
somewhat closer match, but their LF still overpredicts our number counts
and SFR by $\sim2.5$ (Table~\ref{tab:lumlimits}). While the \Lstar\ seems
to reproduce a closer match to the proportion of the SFR we find in our
detected galaxies (compared to \Lstar\ found by \cite{ymf+99}) their value
of \phistar\ seems too large to fit our data.  In fact, this discrepancy is
accounted for by aperture losses. Our fibre size is 1.1\arcsec\ diameter
but Tresse et al. 2002 used a 2\arcsec\ slit, which is better matched to
the average galaxy size. Once we apply aperture corrections (detailed in the
following Section \ref{subsec:apcors}) our result is comparable. 
  
\begin{table*}
\hspace{-1.5cm}
\footnotesize
\caption[Star formation rate densities from various luminosity functions]{\footnotesize Top two rows show our observed numbers and SFRs
  to both the limiting flux in an average fibre and the effective limiting
  flux (per object) when stacking the spectra. The numbers given are for
  the sample of 38 objects, corrected by a factor 0.42 for
  completeness. The second row includes aperture corrections. The lower
  limit on the SFRD is found by dividing the SFR by the total co-moving
  volume surveyed (V$_{cm}=2.2\times10^4\,{\rm Mpc}^3$). We compare our result with various luminosity functions from the literature, determining predicted number counts and SFRs to the same limiting flux, in our surveyed volume. References are: Y99: Yan et al. (1999), T02: Tresse et al. (2002), G95:Gallego et al. (1995), H00: Hopkins et al. (2000; two sets of parameters are given corresponding to upper and lower limits on the \Ha\ luminosity function). Also tabulated are a few test functions used to investigate the effect of varying \Lstar\ and \phistar. Although not a rigorous fit this is a useful comparison and indicates a correction of the order of $15-20\%$ for incompleteness due the luminosity bias. (\Lstar\ is in units of \ergs, \phistar\ in Mpc$^{-3}$.) }
\begin{tabular}{c|ccc|c|c|c|c|c|c|c}
\hline
\hline
Ref. & \multicolumn{3}{|c|}{Luminosity fn params$^{{\mathrm a}}$} &$<z>$ &\multicolumn{2}{c}{limiting flux} & \multicolumn{2}{c}{limiting flux}  & total SFR & SFR density\\
          & log(\Lstar) &  \phistar  & $\alpha$ & &\multicolumn{2}{c}{$f(\Ha)>1.9\times 10^{-16} $} &\multicolumn{2}{c}{$f(\Ha)>3.3\times 10^{-17}$} &\Msol/yr  &\Msol/yr/Mpc$^3$  \\
          &        &   ($\times10^{-3}$)  &            & &\multicolumn{2}{c}{\ergscm} & \multicolumn{2}{c}{\ergscm}  & \\ 
       & &       &   &   & \# /$50.25$\arcmin $^{2}$ & SFR ($>$lim) & \# /$50.25$\arcmin $^{2}$ & SFR ($>$lim)  & & \\
\hline
This paper  & \multicolumn{3}{|c|}{observed}                 & 0.82 & 21 & 167 & 90 & 312 & -- & $>$0.014  \\
This paper  & \multicolumn{3}{|c|}{ with ap. corrections}    & 0.82 & 21 & 283 & 90 & 746 & -- & $>$0.034  \\

Y99  & 42.82 &  1.5 & -1.35 & 1.3    & 83  & 1782  & 257  & 2225  & 2428  & 0.109 \\
T02  & 41.97 &  4.3 & -1.31 & 0.74   & 26  & 261   & 205  & 680   & 922   & 0.041 \\
G95  & 41.56 &  1.6 & -1.3  & 0.0225 &  1  &  9    &  33  & 72   & 133   & 0.006 \\
H00  & 42.88 & 0.77 & -1.6  & 1.3    & 66  & 1274  & 285  & 1789  & 2284  & 0.102 \\
H00 & 43.35 & 0.088 & -1.86 & 1.3    & 35  &  778  & 193  & 1125  & 2324  & 0.104 \\ 
Test & 41.97 & 2.0 & -1.35   & --    & 12  & 121   & 100 & 325 & 456 & 0.02  \\
Test & 41.97 & 1.8 & -1.35   & --    & 11  & 109 & 90 & 293 & 410 & 0.018 \\
Test & 42.00 & 4.0 & -1.35   & --    & 27  & 279 & 212 & 709 & 977 & 0.044 \\  
Test & 42.00 & 3.8 & -1.35   & --    & 26  & 265  & 201 & 674 & 928 & 0.042 \\ 
\hline
\end{tabular}
\begin{list}{}{}
\item[$^{{\mathrm a}}$] converted to our $\Lambda$ CDM cosmology. The
  original luminosity function parameters quoted by the authors are: Y99:
  $L^*=10^{42.85}\,{\rm erg\,s}^{-1}$, $\phistar =
  1.7\times10^{-3}\,{\rm Mpc}^{-3}$, $\alpha=-1.35$ (for
  $H_0=$50\,\kms\,Mpc$^{-1}$, $q_0=0.5$); T02: $\Lstar=10^{42.37}\,{\rm erg\,s}^{-1}$, $\alpha=-1.31$,
  \phistar$=4.07\times10^{-3}$\,Mpc$^{-3}$, 
  ($H_0=$50\,\kms\,Mpc$^{-1}$, $q_0=0.5$) reddening corrected;  G95:$\alpha
  =-1.3$, \Lstar $= 10^{42.15}$\ergs, \phistar = $6.3\times10^{-4}$Mpc$^{-3}$ ($H_0 = 50$\kms\ Mpc$^{-1}$) reddening corrected; H00: \phistar=$3.3\times10^{-4}$,\Lstar=$10^{43.02}$, $\alpha=-1.86$, OR \phistar=$2.9\times10^{-3}$,\Lstar=10$^{42.55}$, $\alpha=-1.6$ ($H_0=75$\kms\ Mpc$^{-1}$,$q_0=0.5$).

\end{list}

\label{tab:lumlimits}
\normalsize
\end{table*}

\subsection{Aperture Corrections}
\label{subsec:apcors}

There is no fool-proof method to calculate aperture corrections for our \Ha\ fluxes. We attempt this using the GOODS $B-$band imaging, which at redshift $z\sim1$ samples the rest-frame UV. We perform photometry on these images using \texttt{phot} in {\sc IRAF}, with the same centres as our fibre positions, and 1\arcsec\ apertures. The aperture correction is then the difference between the resulting 1\arcsec\ aperture magnitudes and the total magnitudes listed in the GOODS v1.0 catalogue\footnote{Publicly available at http://www.stsci.edu/science/goods/}. We then apply the same fractional correction to the \Ha\ flux. This relies on the assumption that the distribution of \Ha\ directly traces the UV, that there is no gradient in reddening across the galaxy. 

\begin{figure}
\centering
\includegraphics[width=8cm]{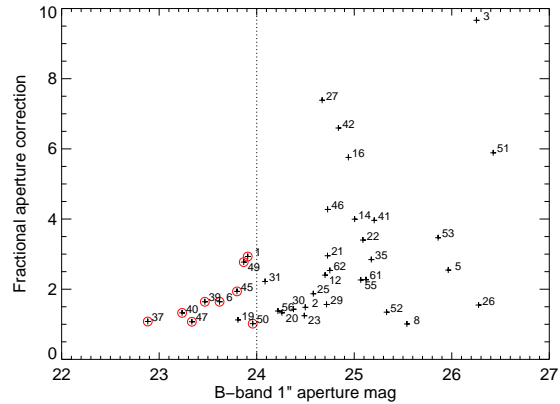}
\caption{$B-$band 1\arcsec\ aperture magnitudes versus the fractional correction necessary to multiply by to obtain total $B-$magnitudes for our observed sample. The galaxies detected in \Ha\ are circled. Note that we detect all but one galaxy brighter than $B=24$. The overlaid object identification numbers allow a comparison with Table~3.6 and Figure~3.9 in the PhD thesis of Doherty (University of Cambridge, 2005), which display the $B-$band (rest-UV) properties of these objects. }
\label{fig:apcors}
\end{figure}

Figure~\ref{fig:apcors} shows the $B-$band 1\arcsec\ aperture magnitudes versus the fractional correction necessary to obtain total $B-$magnitudes for our observed sample, with the galaxies detected in \Ha\  marked with circles. We detect all but one galaxy brighter than $B=24$, indicating that perhaps the reason we don't detect more galaxies in \Ha\ is simply the limit of the instrument sensitivity.

As a consistency check we also looked at the 1\arcsec\ aperture magnitudes
from the GOODS catalogue. However, we have not used these in calculating aperture corrections as our centres are very slightly offset from the GOODS centres, in the most extreme case by $\sim0.5$\arcsec\ for very large/irregular galaxies. This therefore can significantly affect the 1\arcsec\ magnitudes -- we need to compare to the same positioning as our fibres in order to extrapolate the \Ha\ flux correction. It does not, however, affect the total magnitudes and we have therefore taken these directly from the GOODS v1.0 catalogue. 

Applying the measured fractional aperture corrections to the \Ha\ flux measures (or upper limits) leads to an overall aperture correction of a factor of 2.4 for the stack. Adjusting the SFR and SFRD estimates by this amount then gives $746\pm224\msol\,yr^{-1}$ and an SFRD of $0.034\pm0.010\,\Msol\,yr^{-1}$\,Mpc$^{-3}$ (for a co-moving volume of $2.2\times10^4$Mpc$^3$), which is more in line with the measurement of Tresse et al. (2002).
From Table~\ref{tab:lumlimits} it can be deduced that the completeness correction for luminosity bias is most likely of the order of 15$-$20\%, giving $\sim0.04$\,\Msol\,yr$^{-1}$\,Mpc$^{-3}$.

\subsection{Reddening corrections}
\label{subsec:reddening}

\Ha\ emission is more robust to extinction than the UV continuum but not
completely immune. In the following section we firstly compare uncorrected
values of the \Ha\ derived SFRD. However, this assumes that the amount of
dust extinction does not evolve over cosmic time and also that it is
independent of the SFR (i.e. involves the same scaling over all redshift
bins) which is known to be untrue (star formation creates dust therefore
extinction is quite heavily dependent on the amount of ongoing star
formation). The final correction needed therefore, is for dust reddening. 

The usual method for correcting \Ha\ is to compare the observed Balmer
decrement (\Ha /H$\beta$ ratio) with its theoretical value (=2.86;
Osterbrock 1989). We did not observe the H$\beta$ line as our wavelength
coverage was insufficient, we instead work with the UV-continuum from the
$B-$band magnitudes measured above, and use the observed SFR(\Ha):SFR(UV)
ratio to derive an estimate of the extinction using both the Calzetti
(1997) law and a Milky Way law, for comparison with other authors
(Section~\ref{sec:madau-diag}). 

We calculated rest-frame UV (2400\AA) luminosity
densities and corresponding star formation rate using the conversion
given in Kennicutt (1998)\nocite{ken98}:
\begin{equation}
{\rm SFR}(\Msol~{\rm yr}^{-1})=1.4\times10^{-28}L_{\nu}~~~~~({\rm erg~s}^{-1}~{\rm Hz}^{-1})
\label{sfr_uv}
\end{equation}
which assumes continuous star
 formation over $\sim10^8$ years, a flat continuum in the UV (in $f_{\nu}$) and the same IMF as used when deriving SFRs from the \Ha\ flux
 (Equation~\ref{eqn:K98}). 

\begin{figure}
\psfig{figure=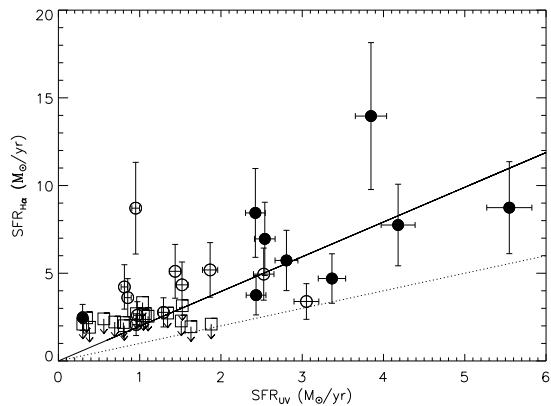,width=8cm} \\
\caption{Comparison of SFRs obtained from UV continuum flux at 2400\AA\
  versus \Ha\ flux for the individual galaxies. The  SFRs derived from UV luminosity
  are consistently underestimated. The filled circles are the robust detections, the open circles are greater than 3$\sigma$ and the squares represent 3$\sigma$ upper limits. The solid line has a gradient of 1.98 and represents the line of best fit to the data (using a least squares fit, through zero). The errors in the SFR(\Ha) values reflect the 30\% error on the flux calibration. We assume $\sim5\%$ error in the UV flux, which is
  dominated by errors in positioning the fibres (i.e. mirrored in the exact
  positioning of the aperture for the $B_{AB}$ magnitude). The dotted line is the line of zero extinction i.e. where SFR(\Ha)$=$SFR(UV).}
\label{fig:sfr_ratio}
\end{figure}

Figure~\ref{fig:sfr_ratio} shows the SFRs for each galaxy calculated using
the UV luminosity density and the \Ha\ flux: those calculated from the UV
luminosity densities are a factor of 1.98 lower on average than those from
\Ha. This is probably due to the differential effect of dust extinction
in the redshift one galaxies between $\lambda_{\rm rest}\approx$ 2400\AA\ and 6563\AA.
This is consistent with results obtained by Glazebrook et al. (1999),
Tresse et al. (2002), and Yan et al. (1999)\nocite{ymf+99} who all
find SFR(\Ha)/SFR(UV) ratios of around $2-3$.

There is no uniform correction that can be applied to the SFR estimated
from UV flux in order to derive the true star formation rate: the amount of extinction varies
in each object due to inherent differences in physical properties of
the galaxies (e.g. Sullivan et al. 2001\nocite{smc+01}). The errors on
our \Ha\ fluxes are not great enough to account for the
scatter in Figure~\ref{fig:sfr_ratio}, which instead is attributable to different dust extinctions in our galaxy
sample. However, we can estimate the true star formation rate for the
sample, using the average SFR(\Ha)/SFR(UV) ratio if we assume that
the Kennicutt (1998) relations for \Ha\ and UV SFRs should yield the same
intrinsic SFR in the absence of extinction.

Table~\ref{tab:Av} gives the values of the extinctions in magnitudes $A_v$\footnote{Note that the correction from $A_v$ to $A_{\Ha}$
  is small, but non-negligible. $A_{\Ha}=0.82A_v$ for a Milky Way law and
  $0.85A_v$ for a Calzetti law.}, $A_{\Ha}$, the colour excess $E(B-V)$ and the conversion
factor to convert SFR(\Ha) to a total SFR, for the Milky Way (Seaton 1979\nocite{sea79}) and Calzetti 
laws (Calzetti 1997), which yield our observed ratio SFR(\Ha)/SFR(UV)=1.98.
For the purposes of comparison in the Madau-Lilly diagram in the following section, we show reddening corrections using both extinction laws.

\begin{table}
\centering
\caption[Comparison of Milky Way and Calzetti extinction laws]{$E(B-V)$,
  $A_v$,  $A_{\Ha}$ and correction factor to convert SFR(\Ha) to a total SFR, for the Milky Way and Calzetti laws}
\begin{tabular}{c|cccc}
\hline
 & $E(B-V)$ & $A_v$ & $A_{\Ha}$ & correction factor \\
\hline
Milky Way &  0.156 & 0.5 & 0.41 & 1.46 \\ 
Calzetti  & 0.16 & 0.78 & 0.65  & 1.85 \\
\hline
\end{tabular}
\label{tab:Av}
\end{table}

\section{Madau-Lilly Diagram: evolution of the star formation rate density}
\label{sec:madau-diag}
\begin{figure}
\centering
\epsfig{figure=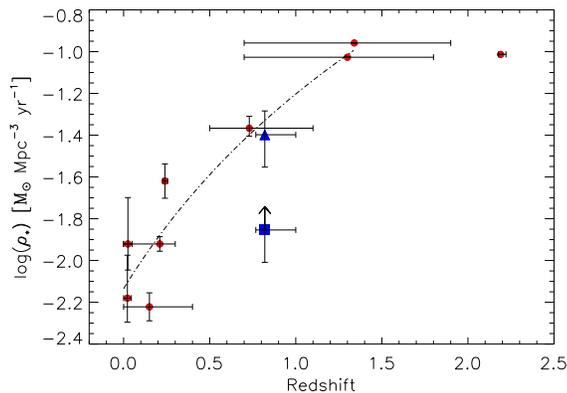,width=8cm}
\caption[Evolution of the star formation rate density]{\small{SFRDs determined from \Ha\ measurements only, with no
    reddening corrections. Circles are points taken from the
    literature, converted to our cosmology (see
    Table~\ref{tab:madau-values}). Overlaid is our lower limit (square) to the SFRD, and our estimate including luminosity bias and
    aperture corrections (triangle). The dashed line represents a
    $(1+z)^{3.1}$ power law. Our corrected point is consistent
    with previous determinations of evolution in the SFRD according to
    $(z+1)^4$ (in an $H_0=50$\,\kms~Mpc$^{-1}$, $q_0=0.5$ cosmology) from $z=0-1$ (e.g. Tresse et al. 2002 and refs therein). }}
\label{fig:mymadau-diag}
\end{figure}

Having finally derived a global SFRD for the redshift range $0.768<z<1.0$,
corrected for incompleteness, fibre aperture losses and with  an average
reddening correction, we turn now to placing this measurement in the context
of other work in the field, to address the question of evolution of the SFR
over cosmic time. As discussed in the Introduction, the past decade has
seen many attempts to correct and calibrate various SFR indicators, in an
attempt to put measurements at different cosmic epochs on the same scale
(the `Madau-Lilly diagram') and by so doing trace the time evolution of the
global star formation rate density. We focus on comparing our
result only to other \Ha\  measurements in the literature, in order to take one consistent indicator across all redshift bins. With current infra-red
technology \Ha\ can be traced out to redshift $z\sim2.5$ ($K-$band) --
i.e. $\sim80\%$ of the age of the universe. The
compilation of \Ha\ SFR measurements from the literature is given in
Table~\ref{tab:madau-values}, converted to our cosmology
(H$_0$=70\,km\,s$^{-1}$\,Mpc$^{-1}$, $\Omega_M=0.3$,
$\Omega_{\Lambda}=0.7$). These are then used to plot the evolution of the
SFRD (uncorrected for reddening) in Figure~\ref{fig:mymadau-diag}. Some
authors have corrected their data for reddening and for the purposes of
comparison with our result, following Tresse et al. (2002), we artificially
`redden' their result using the canonical $A_v=1$, (a factor 2.02 at \Ha). We overplot two values from this work -- the first, lower point is our measured SFRD with no corrections (other than for the completeness of the observed sample) which represents a firm lower limit on the SFRD in the redshift range 0.768--1.0 . The upper value includes corrections for incompleteness due to luminosity biasing and aperture corrections. The error bars reflect the uncertainty in the flux calibration, which is $\sim30\%$. 

There may be a small amount of contamination from Active Galactic Nuclei
(AGN). We have not corrected for this, both as the effect is very minimal
(well within the 30\% systematic errors) and also as the other SFR
estimates on the Madau-Lilly diagram in Figure~\ref{fig:mymadau-diag} have not
been corrected. Pascual et al. (2001) estimate that at low redshift AGN
contribute about 10\% by number and $\sim15\%$ of the \Ha\ luminosity
density. Matching our target list of galaxies observed with CIRPASS to the
Chandra 2Ms X--ray catalogue (Alexander et al. 2003\nocite{abb+03}), we
find six sources out of our 62 observed which are likely AGN. Only three of
these are included in the stacked spectrum from which we estimate the SFRD,
(the others were excluded due to \Ha\ falling in the water absorption
region of the spectrum) the combined SFR of these three sources as inferred from their \Ha\ flux is 12.5\Msol\,yr$^{-1}$, or 9.5\% of the total for our observed sample. This constitutes an {\it upper} limit to the AGN contamination as not all of the \Ha\ flux is produced by the AGN, it is likely a mixture from both the AGN and star formation.

\begin{figure}
\centering
\epsfig{figure=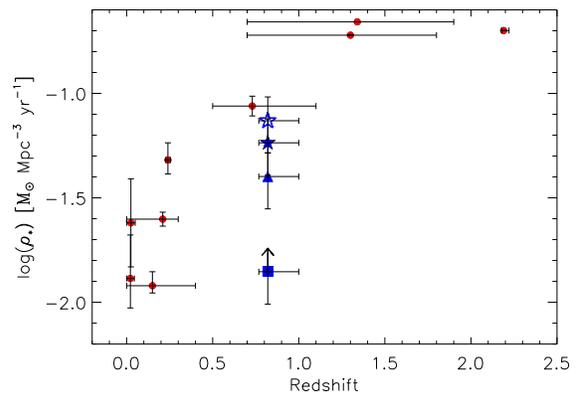,width=8cm}
\caption[Evolution of the reddening corrected star formation rate density]{\small{ As for Figure~\ref{fig:mymadau-diag} but with reddening
    corrections. We have used the authors own reddening corrections where
    applicable. For the three highest redshift points, the authors have not
    given reddening corrections and we use a correction $A_v=1.0$, or a
    factor 2.02 at \Ha, following Tresse et al. (2002). Our uncorrected data points are
    shown with a square and triangle, as above. Reddening
    corrected points are shown for the Milky Way law (filled star) and a Calzetti
    law (open star).}} 
\label{fig:mymadau-diag-red}
\end{figure}

As some authors have corrected their data for reddening, we make a separate
comparison to the reddening corrected SFRDs in the literature in
Figure~\ref{fig:mymadau-diag-red}. We use the average reddening on our data
as derived above -- a factor 1.46 for a Milky Way law or 1.85 for a
Calzetti law, to compare our result. As the authors who derived the three
highest redshift data points (Yan et al. 1999, Hopkins et al. 2000 and
Moorwood et al. 2000) have not applied reddening corrections we adjust their
data points by $A_v=1$ (or a factor 2.02, following Tresse et al. 2002). 

Our SFRD at $0.768<z<1.0$ is consistent with the value derived by Tresse et al. (2002) over a similar redshift range, although our data do not support quite the same strength of reddening as they invoke. 
Our result is still a factor of $\sim$2.5 lower than the results of \cite{ymf+99} and \cite{hcs00}. However, these works span a much greater redshift range, probing out to $z\sim1.8$. 

The evolution in luminosity density, or star formation rate density, of the
Universe is often fit by a power law of the form $(1+z)^n$. Fitting a least-squares
line in the log(SFRD)$-$log(1$+z$) plane yields
$(1+z)^{3.1}$ (Figure~\ref{fig:mymadau-diag}), which is strong evolution from $z=1$ to the $z=0$ universe,
and comparable to previous work. Tresse et al. (2002) found evolution in
the star formation rate density as $n=4.1$ from \Ha\ (for an $H_0=50$\kms\ Mpc$^{-1}$,
$q_0=0.5$ cosmology), which was in agreement with the evolution in UV
(2800\AA) luminosity density in the CFRS (Lilly et al. 1996). In fact, the
change in cosmology softens the power law. Without refitting their data, using Gallego's value as a zero point and comparing to that of
Tresse, we find that the cosmology conversion would change the power law to $n\approx  3.5$. We therefore find a power law slightly shallower than those previous works, but our data nevertheless supports a steep decline in the star formation rate from $z=1$ to $z=0$.

\begin{table*}
\footnotesize
\hspace{-1cm}
\caption{Values in the literature for \Ha\ luminosity densities (L(\Ha) in \ergs Mpc$^{-3}$) or
  \Ha\ derived star formation rate densities. Values of L(\Ha) have been
  converted to our cosmology, $H_0=70$\,\kms~Mpc$^{-1}$, $\Lambda= 0.7$, $\Omega =0.3$, using the mean redshift $<z>$.
  Where the authors have applied a reddening correction to their data, we
  use a value of A$_v$=1 mag to `de-redden' their result for the no
  reddening column. Conversely, if no reddening correction has been applied
  in the original work, we redden the result by A$_v$=1 mag to obtain the
  values listed under the reddening column. Star formation rate densities
  have been derived using the converted L(\Ha) value and the K98
  conversion. References are G95:\cite{gzar95} , PG03:\cite{pzg+03} ,
  S00:\cite{ste+00} ,T98:\cite{tm98} , P01:\cite{pgaz01} , T02:\cite{tmlc02}
  , D06: this paper, H00:\cite{hcs00} , Y99:\cite{ymf+99} , M00:\cite{mwco00} .  }
\begin{tabular}{cllllllll}                   
\hline
\hline
    &       &          &             &                &           \multicolumn{2}{c}{reddening}                 &     \multicolumn{2}{c}{no reddening}       \\         
ref & $<z>$ & redshift & original    &  original     & converted  & SFRD                  & converted   & SFRD    \\
    &       & range    & cosmology   & log(L$_{\Ha}$)  &  log(L$_{\Ha}$)     &                        & log(L$_{\Ha}$)   &          \\
\hline  

G95 & 0.022 & $z\lesssim0.045$ &$H_0=50$&39.10$\pm$0.2$^{{\mathrm a}}$  & 39.23$\pm$0.2  & 0.013$^{+0.008}_{-0.005}$ & 38.92$\pm$0.2 & 0.0066$^{+0.004}_{-0.002}$ \\
PG03 & $\sim0.025$ & $z\lesssim0.05$ & $\Lambda$CDM$^{{\mathrm b}}$ & 39.5 $\pm$0.2 $^{{\mathrm a}}$ & 39.5  $\pm$0.2  & 0.024$\pm$0.015  & 39.195$\pm$0.2 & $0.012^{+0.008}_{-0.004}$ \\
S00 &  0.15  & $0<z<0.4$       & $H_0=100,q_0=0.5$& 39.49$\pm$0.06 $^{{\mathrm a}}$ & 39.19 $\pm$0.06 & $0.012^{+0.002}_{-0.001}$ & 38.89 $\pm$0.06 & 0.006 $\pm$0.001 \\
T98 & 0.21  & $0<z\leq0.3$     & $H_0=50,q_0=0.5$ & 39.44$\pm$0.04 $^{{\mathrm a}}$ & 39.50 $\pm$0.04 & 0.025 $\pm$0.002    & 39.20 $\pm$0.04& 0.012 $\pm$0.001\\
P01 &  0.24 & $ 0.228<z<0.255$ & $H_0=50,q_0=0.5$ & 39.73$\pm$0.09 $^{{\mathrm a}}$ & 39.79 $\pm$0.09 & 0.048$^{+0.01}_{-0.008}$  & 39.48 $\pm$0.09 & 0.024 $\pm$0.005 \\
T02 &0.73&$0.5<z<1.1$&$H_0=50,q_0=0.5$& 40.10$\pm$0.05$^{{\mathrm a}}$&40.04$\pm$0.05&0.087$\pm$0.01&39.74$\pm$0.05&0.043$^{+0.006}_{-0.004}$ \\
D06 & 0.82 & $0.77<z<1.0$ &   $\Lambda$CDM$^{{\mathrm b}}$ & $39.70^{+0.11}_{-0.15}$ & $39.97^{+0.11}_{-0.15}$ & 0.074$\pm0.022$ & $39.70^{+0.11}_{-0.15}$ & 0.04$\pm$0.012 \\
H00 &1.3& $0.7<z<1.8$&$H_0=75,q_0=0.5$& 40.32 & 40.38         & 0.19          & 40.08           & 0.094 $^{{\mathrm d}}$          \\
Y99 &1.34 & $0.7<z<1.9 $& $H_0=50,q_0=0.5$ &  40.22         & 40.45         &  0.22               &   40.15  & 0.11 $^{{\mathrm d}}$               \\
M00 & 2.19  & $2.178<z<2.221$  & $H_0=50,q_0=0.5$ & 40.19 $^{{\mathrm c}}$       & 40.39         & 0.20              & 40.09 & 0.097 $^{{\mathrm d}}$\\

\hline
\end{tabular}                                                                                                                           
\footnotesize
\begin{list}{}{}
\item[$^{{\mathrm a}}$] Authors have applied a reddening correction to their data before deriving L(\Ha). To compare, we `re-redden' L(\Ha) by the `canonical' value of A$_v$=1 mag following Tresse et al. (2002) to derive value for the no reddening column.  
\item[$^{{\mathrm b}}$] $H_0=70$\,\kms~Mpc$^{-1}$, $\Lambda= 0.7$, $\Omega =0.3$
\item[$^{{\mathrm c}}$] Authors use the luminosity function of Yan et al. (1999) to derive this value
\item[$^{{\mathrm d}}$] Authors do not quote errors in the original work

\end{list}
\normalsize
\label{tab:madau-values}
\end{table*}

\section{Summary}

We have carried out an \Ha\ survey of redshift one galaxies in the HDF-N,
using multi-object near infrared fibre spectroscopy. We found 9 robust
detections ($>5\sigma$), of which the brightest are presented in Doherty et al.\, (2004). The real power of this
technique lies in stacking the signal to obtain a star formation rate
estimate for the sample as a whole.

We have derived a {\em lower} limit to the SFR of 312 \msol\,yr$^{-1}$ (corrected for incompleteness due to observing a random sub-sample of all possible targets), or an SFRD of 0.014 \Msol\,yr$^{-1}$\,Mpc$^{-3}$ . Correcting this for volume incompleteness due to the luminosity bias effect and aperture losses gives an SFRD of 0.04 \Msol\,yr$^{-1}$\,Mpc$^{-3}$. This is consistent with the work of Tresse et al. (2002) over a similar redshift range. Yan et al. (1999) and Hopkins et al. (2000)  find a result $\sim2.5$ times higher, from space-based grism spectroscopy, but their work covers a much wider redshift range ($z\sim0.7-1.9$), with a mean $<z>$=1.3 and we therefore do not expect it to be representative of $z=0.8$ if the \Ha\ luminosity density continues to rise at $z>1$. 

Our SFRD estimate is consistent with steep evolution from $z=0$ to $z=1$, described by a power law $(1+z)^{3.1}$. 

We finally derived an average reddening correction of $A_v$=0.5--0.78
magnitudes, or a factor 1.46--1.85 at \Ha\ (depending on the extinction
law), and place this in context with reddening corrected values from the
literature (Figure~\ref{fig:mymadau-diag-red}). However, this may be an
underestimate of the reddening as it was derived using a comparison to the
UV luminosities, and the UV luminosity can continue to evolve even after
star formation has ceased leading to an additional uncertainty in the
SFR$_{\rm UV}$. A caveat to this work which should be noted is that the result is obtained for one small field of view and thus is cosmic variance dependent. 

We have demonstrated the first successful application of multi-object fibre spectroscopy to observe high redshift galaxies. In spite of the many difficulties involved in searching for such weak signals, this approach constitutes a powerful technique and the success of our multi-object spectroscopic survey bodes well for larger surveys with future instruments such as FMOS on Subaru \citep{ldh+03,kmo+03} and EMIR on GTC (Garzon et al. 2004).

\subsection*{ACKNOWLEDGMENTS}

We thank the anonymous referee for helpful comments which have improved the paper. This paper is based on observations obtained at the William Herschel
Telescope, which is operated by the Isaac Newton Group on behalf of
the UK Particle Physics and Astronomy Research Council. We thank the
WHT staff, in particular Danny Lennon, Kevin Dee, Rene Ruten, Juerg Rey and Carlos Martin,
for their help and support in enabling CIRPASS to be used as a visitor
instrument.  Simon Hodgkin, Elizabeth Stanway, Emily MacDonald and Paul Allen 
assisted in obtaining the observations.  CIRPASS was built by the
instrumentation group of the Institute of Astronomy in Cambridge, UK.
We thank the Raymond and Beverly Sackler Foundation and PPARC
for funding this project. We are indebted to Dave King, Jim Pritchard,
Anthony Horton \& Steve Medlen for contributing their instrument
expertise. We are grateful to Steve Lee and Stuart Ryder at AAO for
assistance in designing the fibre plug plates, and thank the AAO for
the use of the FOCAP fibre unit. The optimal extraction software for
this fibre spectroscopy was written by Rachel Johnson, Rob Sharp and
Andrew Dean.  This research is also partially based on observations
with the NASA/ESA {\sl Hubble Space Telescope}, obtained at the Space
Telescope Science Institute (STScI), which is operated by AURA under
NASA contract NAS 5-26555. These observations are associated with
proposals \#9425\,\&\,9583 (the GOODS public imaging survey). We also
used results from the Caltech Faint Galaxy Redshift Survey in the
HDF-N, and thank Judy Cohen and colleagues for making these catalogues
publically available. MD is grateful for support from the Fellowship
Fund Branch of AFUW Qld Inc., the Isaac Newton Studentship, the
Cambridge Commonwealth Trust and the University of Sydney.

\bibliographystyle{mn2e}
\bibliography{refs}

\section*{Appendix}
\label{app:A_lumfunc}
  The conversion of an \Ha\ luminosity function from the original cosmology
  to that assumed throughout this paper is as follows (following Hopkins
  2004\nocite{hop04}). For a flat universe, the co-moving volume in a certain redshift rang
e
  $z_1<z<z_2$ is proportional to the difference of the
  co-moving distances ($D_c$) cubed, i.e. $V_{cm}\propto
  D_c(z_2)^3-D_c(z_1)^3$ and luminosity is proportional to the co-moving
  distance squared, $L\propto D_C^2$. To convert a luminosity function we
  convert the luminosity and number density (i.e. volume) parameters
  separately, by evaluating those expressions at the appropriate redshift,
  for each cosmology. The ratio gives the conversion factor, i.e. \Lstar\ is
  multiplied by the ratio of the squares of the co-moving distances and
  \phistar\ is divided by the ratio of the co-moving volume shells (between
  $z_1$ and $z_2$). So to
  convert a luminosity function parametrised as \Lstar, \phistar, $\alpha$
  in a cosmology where the co-moving distance is $D_c$, to $\Lstar^{\prime}$,
  $\phistar^{\prime}$, $\alpha^{\prime}$ with $D_c^{\prime}$, we have
\begin{equation}
\Lstar^{\prime}=\Lstar \times D_c^{\prime 2}/D_c^2,
\end{equation}
and
\begin{equation}
\phistar^{\prime}= \phistar \times \frac{D_c(z_2)^3-D_c(z_1)^3}{D_c(z_2)^{\prime 3}-D_c(z_1)^{\prime 3}}
\end{equation}

This conversion relies on the assumption that all galaxies lie at the
central redshift, a more rigourous method would be to correct the original
data points and refit the Schechter function.
However, the likely error associated with employing this approximation
method is $\sim10\%$ (Hopkins 2004).

\label{lastpage}

\end{document}